# Wave nature-based nonlocal correlation via projection measurements between space-like separated interferometric systems

Byoung S. Ham
School of Electrical Engineering and Computer Science, Gwangju Institute of Science and Technology
123 Chumcangwagi-ro, Buk-gu, Gwangju 61005, S. Korea
(Submitted on Feb. 23, 2022, bham@gist.ac.kr)

**Abstract:**
Indistinguishability in quantum mechanics is an essential concept to understanding "mysterious" quantum features such as self-interference of a single photon and two-photon nonlocal correlation. Delayed-choice experiments are for the cause-effect violation via post-measurements of photons in an interferometric system. Recently, a macroscopic version of the delayed-choice experiments has been demonstrated using the classical means of Poisson distributed photons in a noninterfering Mach-Zehnder interferometer (NMZI). Here, a coherence version of the nonlocal correlation is presented using the wave nature of photons in space-like separated orthogonally polarized NMZIs. For this, polarization-basis projections of the pair of NMZI output photons onto a set of polarizers are measured coincidently, where each output photon comprises orthogonal polarization bases. Because of the coherence feature of the output photons via coincidence detection, a phase sensitive nonlocal correlation is achieved. As a result, indistinguishability limited to a single or entangled photon pairs needs to be reconsidered with basis (polarization) randomness for the nonlocal correlation.

**Introduction**

In quantum mechanics, measurements affect the original quantum state [1,2]. Measurement-based quantum mystery has been studied for various delayed-choice experiments over the last several decades [3-12]. Not only a single photon-based quantum superposition [4-6,8], but also two photon-based nonlocal correlation [10-13] has been intensively studied for the mysterious quantum feature violating causality. Potential quantum loopholes in such a quantum system have also been closed for detection [13,16], locality [14], sampling [13,15,16], and free will [17] parameters. The nonlocal correlation between space-like separated local measurements results in nonlocal realism of quantum mechanics [18]. Classical realism is for an independent reality not influenced by the action of measurements or observations as claimed by Einstein and his colleagues [19]. Nonlocal realism defies the classical reality via quantum features of "spooky action at a distance" [20]. This nonlocal correlation violating local realism is the unique feature of quantum mechanics that cannot be explained or achieved by any classical means. To satisfy the nonlocal correlation, it is commonly accepted that target photons must be nonclassical [13-23].

Recently, the quantum feature of Franson-type nonlocal correlation [24,25] has been newly interpreted with the wave nature of photons, i.e., coherence optics, where the phase-determined nonlocal correlation should be provided by local realism of coherence optics [26]. In other words, each entangled photon entering a noninterfering MZI of the Franson scheme must satisfy self-interference even under the condition of overall (effective) decoherence, where the coherence of each entangled photon pair relies on the pump photon due to phase matching condition of $\chi^{(2)}$ nonlinear optics. Thus, each locally measured photon should show a typical interference fringe, i.e., self-interference if the pump laser is quite narrow compared with that of entangled photons. The wide bandwidth-distributed photon ensemble generated from a spontaneous parametric down conversion process (SPDC) washes out this coherence feature in each photon pair, resulting in local randomness [27]. Here, the key factor is that the bandwidth of an entangled photon ensemble is much broader than that of each photon [26,27]. With the proper design of an asymmetric MZI, the two features of single-photon coherence and ensemble decoherence can be simultaneously provided [25,26]. Thus, the Franson-type nonlocal correlation based on each photon pair does not have to be affected by the ensemble decoherence if coincidence detection is involved [26].

In this paper, a contradictory phenomenon of nonlocal correlation to the conventional understanding of quantum mechanics is presented using the classical means of attenuated laser light to speculate the quantum feature.

The conventional understanding of quantum mechanics is in the microscopic regime affected by measurements. Thus, the wave-particle duality of a single photon can never be determined a priori but decided a posteriori by measurements [3-12]. Because an interference fringe of MZI cannot distinguish between single and ensemble photons [28], the delayed-choice experiments in an MZI can be extended into a macroscopic version [29]. Here, a coherence version of nonlocal correlation is proposed in a Franson-like scheme using quantum projection measurements of the MZI output photons. Unlike the original Franson scheme based on phase bases [25], polarization bases of a photon are used for the nonlocal correlation between paired noninterfering MZIs, where the quantum projection of the MZI output photons plays a key role. For this, MZI comprises a beam splitter (BS) and a polarizing BS (PBS), satisfying the Fresnel-Arago law and the pre-set particle nature of a photon [30]. Instead of using entangled photon pairs as in conventional nonlocal correlation measurements [13-23], a pair of doubly bunched photons from an attenuated laser is used for the cross-correlation measurements under coincidence detection. Thus, any existence of nonlocal feature between the paired photons should witness the macroscopic nonlocal correlation. To understand this macroscopic nonlocal correlation, a coherence feature between paired photons is newly interpreted for the basis randomness, where the basis randomness has no discrepancy between a single and ensemble entities, resulting in a macroscopic understanding on quantum mechanics.

**Results**

*Coherent photon-based nonlocal correlation*

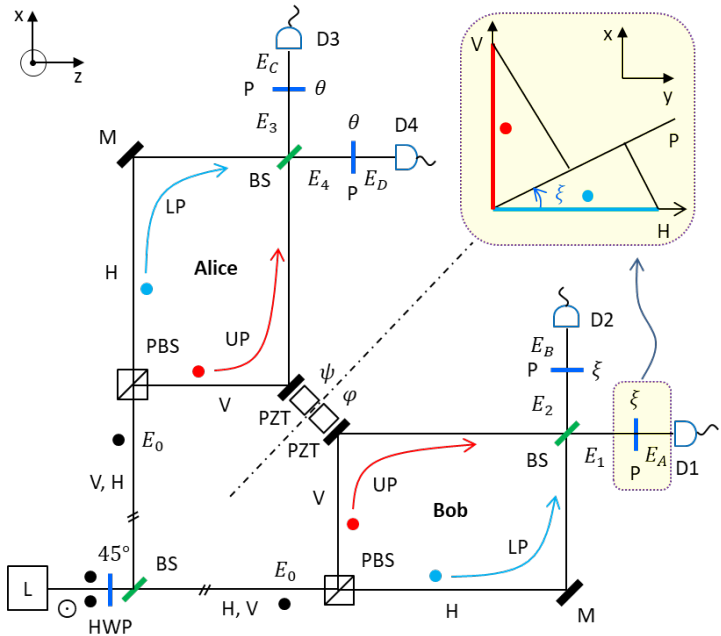

Fig. 1. Schematic of coherence quantum correlation. (Inset) Polarization basis projection onto a polarizer. L: laser, HWP: half-wave plate, BS: beam splitter, PBS: polarizing BS, H (V): horizontal (vertical) polarization, M: mirror, PZT: piezo-electric transducer, P: polarizer, D: single photon detector. The black dot indicates a pair of single photons whose initial polarization is vertical, while the colored dots indicate split photons in an equal chance in both paths.

Figure 1 shows schematic of the coherent photon-based nonlocal correlation between noninterfering MZIs comprising PBS and BS for orthonormal bases of vertical (V) and horizontal (H) polarizations of a photon. Thus, the photon characteristic in a noninterfering MZI is predetermined by the particle nature according to the wave-particle duality in quantum mechanics. For the quantum mystery of "delayed-choice," a polarizer (P) is added for the output photon's projection measurements in the polarization bases. Due to independence between the polarizer and MZI, any violation of the causality witnesses the "delayed choice," resulting in the fundamental quantum feature [3-13]. Such violations have been demonstrated recently for both single photons and continuous wave (cw) light [29].

Regarding the light source in Fig. 1, only doubly bunched photon pairs are selected via coincidence measurements. Although such a choice of doubly bunched photon pairs represents sub-Poisson statistics by eliminating both vacuum and single photon states via coincidence detection, the present nonlocal correlation can also be applied for a classical regime of coherence optics. The input field $E_0$ denotes an amplitude of each photon in the doubly-bunched pair, where the mean photon number of all pairs is set to be extremely low, satisfying the incoherence condition between consecutive photons. For this, the photon-to-photon distance in each MZI must be much longer than the laser's coherence length. Notably, the coherence of an ensemble photons is the same as that of each photon [31]. This the fundamental difference from the SPDC generated entangled photons [26]. Unlike coherent photons, SPDC-based entangled photons governed by nonlinear optics are ultrawide in spectral bandwidth, whereas the bandwidth of each entangled photon pair governed by a pump laser can be much narrower [25]. The different features of photon characteristics are the major requirements for the Franson-type nonlocal correlation [26].

By the action of the polarizer P in Fig. 1, the polarization bases of the MZI output photons are projected onto the polarizer. The pre-set incoherence feature of each MZI output photon become coherent by the projection process, satisfying the delayed-choice experiments [29]. Thus, the local measurements in each MZI are not random anymore. To circumvent this local fringe, two measurement results by a set of polarizer's rotation angles can be arithmetically summed (see Analysis). This addition does not violate quantum mechanics because of linear superposition, and it does not affect nonlocal measurements because of independence between local measurements. This action of projection measurements of the pre-set MZI output photons is the physical origin of the macroscopic quantum feature in the present coherence scheme of nonlocal correlation.

*Analysis*

According to coherence optics for a doubly bunched photon pair in Fig. 1 (see two black dots), each input photon's amplitude for Alice or Bob is denoted by $E_0$. For the noninterfering MZIs, the following matrix representations are achieved:

$$\begin{bmatrix} E_1 \\ E_2 \end{bmatrix} = \frac{E_0}{2} \begin{bmatrix} H - Ve^{i\varphi} \\ i(H + Ve^{i\varphi}) \end{bmatrix}, \tag{1}$$

$$\begin{bmatrix} E_3 \\ E_4 \end{bmatrix} = \frac{iE_0}{2} e^{i\eta} \begin{bmatrix} H - Ve^{i\psi} \\ i(H + Ve^{i\psi}) \end{bmatrix}, \tag{2}$$

where H and V indicate amplitudes of polarization unit vectors. The phase $e^{i\eta}$ in equation (2) is due to the path length difference from L to each MZI. Using an inserted polarizer (P) in each output photon path of MZIs, the output fields of equations (1) and (2) are modified to:

$$E_A = \frac{E_0}{2}(cos\xi - sin\xi e^{i\varphi}) \tag{3}$$

$$E_B = \frac{iE_0}{2}(cos\xi + sin\xi e^{i\varphi}), \tag{4}$$

$$E_C = \frac{E_0}{2} e^{i\eta}(cos\theta - sin\theta e^{i\psi}) \tag{5}$$

$$E_D = \frac{iE_0}{2} e^{i\eta}(cos\theta + sin\theta e^{i\psi}), \tag{6}$$

where the rotation angles of Ps in both parties are distinguished by ξ and θ. Here, the positive rotation angle of ξ and θ is defined by a counterclockwise direction from the horizontal axis (y) of the photon propagation vector (z). The MZI phase control is denoted by φ and ψ. The projection results of the output photons on each polarizer are H → cosζ and V → sinζ, where ζ = {ξ, θ} (see the Inset of Fig. 1). For the negative rotation of P, however, the projections become H → cosζ and V → −sinζ. The polarization basis projection onto P represents the control of post-measurements via the filtering process of the orthonormal polarization bases. This filtering process is a typical method of coincidence detection in quantum technologies. As a result, each input photon of each MZI in Fig. 1 has a wave nature, which is contradictory to the pre-set photon characteristics of the particle nature. This is a violation of the causality in the delayed-choice experiments, where the photon's nature is post-determined by measurements [29].

*Synchronized case*
For further analyses, we first consider a synchronized case with $\psi = \varphi$ and $\xi = \theta$. As shown by $\zeta$ for the synchronized rotation angle of the polarizers in both MZIs in Fig. 1, we denote the synchronized phase as $\rho$ for $\varphi$ and $\psi$. The final intensities measured by corresponding four detectors are then:

$$I_A = I_C = \frac{I_0}{4}(1 - sin2\zeta cos\rho), \quad (7)$$

$$I_B = I_D = \frac{I_0}{4}(1 + sin2\zeta cos\rho), \quad (8)$$

where the global phase $e^{i\eta}$ does not affect the measurements ($I_C$; $I_D$). For the negative $\zeta$, the sign in equations (7) and (8) are reversed. For $\zeta_\pm = \pm\frac{\pi}{4}$ ($\pm 45°$), i.e., random basis projections, the normalized mean nonlocal correlation, $\langle R_{AD}\rangle \left(=\frac{\langle I_A \cdot I_D\rangle}{\langle I_A\rangle\langle I_D\rangle}\right)$, between two remotely separated detectors D1 and D4 is as follows:

$$R_{AD}(\zeta_\pm; \rho) = sin^2(\rho). \quad (9)$$

Equation (9) is regardless of the $\zeta$ basis, where $\zeta \in \{-45°, +45°\}$ as shown in Figs. 2a and 2b. In Fig. 2a, the normalized $\langle R_{AD}\rangle$ is calculated, where both local intensities are shown in Figs. 2c and 2d, respectively. In Fig. 2b, the $\zeta_\pm$-dependent intensity swapping for $I_A$ is shown by the red and blue curves, while $\langle R_{AD}(\rho, \zeta_+)\rangle = \langle R_{AD}(\rho, \zeta_-)\rangle$ (see the dotted curve). Thus, these specific rotation angles $\zeta_\pm$ of the P demonstrates orthogonality in the projection measurements in both parties, representing the polarizer's bases. The local randomness for each detector in each party can be accomplished by summation for $\zeta_\pm$, resulting in $\langle \bar{I}_j\rangle = \frac{\langle I_0\rangle}{2}$, where j=A, B, C, and D. As shown in Fig. 2b, however, this local basis sum does not affect the nonlocal correlation $\langle R_{AD}\rangle$.

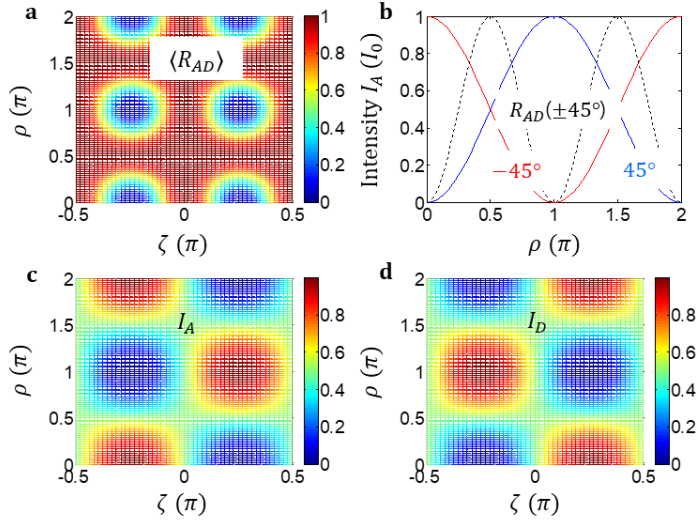

Fig. 2. Numerical calculations for a synchronized case. **a**, $R_{AD}(\rho, \zeta)$. **b**, $I_A(\rho, \zeta_\pm)$. **c**, $I_A(\rho, \zeta)$. **d**, $I_B(\rho, \zeta)$. $\rho = \psi = \varphi$, $\zeta = \xi = \theta$. In a, $R_{AD} = R_{BC}$.

*Nonlocal cross correlation*
Compared with the synchronized nonlocal correlation in equation (9) between two local measurements in space-like separated parties, the intensity correlation between them for $\zeta_\pm$ of Ps is analyzed in Fig. 3. The benefit of energy-phase relation, i.e., cross correlation offers a much more sensitive resolution to nonlocal correlation. Due to the coherence feature, ensemble measurements are advantageous for a single-shot enhanced signal-to-noise ratio in sensitivity. The thumb rule of quantum mechanics is that a photon never interferes with others as claimed by Dirac [34]. Thus, the macroscopic interference in an MZI is actually rooted in the self-interference of a single photon [28].

This fact has been intensively studied over the last decade in Born rule tests using a multi-slit interferometric system [35,36]. Thus, the wave nature-based quantum correlation in Fig. 1 is closely related with the cross-correlation between space-like separated individual measurements.

Figure 3 shows the normalized cross-correlation $R_{AD}$, where equations (7) and (8) are used for a fixed P at $\zeta_\pm$:
$$R_{AD}(\varphi,\psi) = (1 \mp \cos\varphi)(1 \pm \cos\psi). \quad (13)$$
Both Figs. 3a and 3b clearly show the cross-correlation between two remotely separated local measurements, where strong correlation exists between $\varphi$ and $\psi$. The relation between $R_{AD}$ and $R_{BC}$ is anti-correlation in the phase bases $(0; \pi)$ of MZIs. Figures 3c and 3d show details of local measurements of the anti-correlation to support Figs. 2a and 2b, respectively. This phase correlation between remote parties can be applied to the coherence-quantum key distribution (discussed elsewhere).

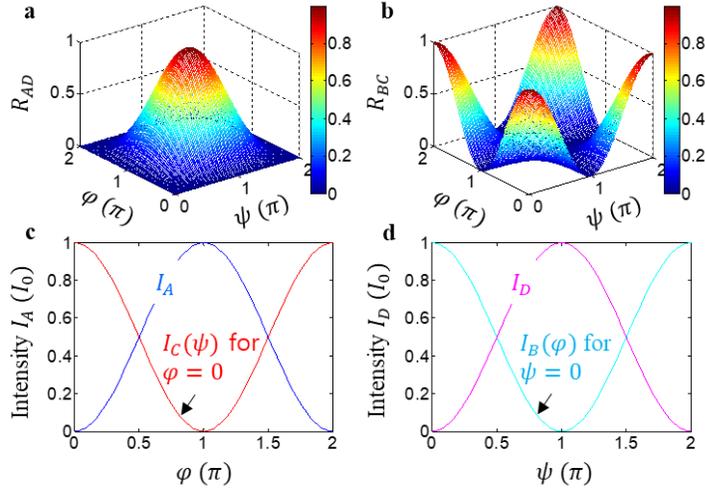

Fig. 3. Numerical calculations for cross-correlation. $\xi = \theta = \frac{\pi}{4}$.

Figure 4 shows normalized cross-correlations, $R_{AD}$ and $R_{BC}$, as a function of the projection angles $\xi$ and $\theta$, respectively. A general relation of $R_{ij}$ is obtained from equations (7) and (8):
$$R_{ij}(\varphi,\psi,\xi,\theta) = (1 - \sin 2\xi \cos\varphi)(1 + \sin 2\theta \cos\psi), \quad (13)$$
where the subscripts $i$ and $j$ indicate local detectors in both parties. The cross-correlation $R_{AD}(\varphi,\xi)$ for $\psi = 0$ and $\theta = \frac{\pi}{4}$ is shown in Figs. 4a and 4b, where, $R_{AD}(\varphi,\xi) = (1 - \sin 2\xi \cos\varphi)$. Figure 4c and 4d represent $R_{BC}(\psi,\theta)$ for $\varphi = 0$ and $\xi = \frac{\pi}{4}$. As shown in Fig. 4, the cross-correlation between remotely separated local measurements exists between $\varphi$ and $\xi$ for fixed $\psi$ and $\theta$, and vice versa. For the corresponding polarizer's bases $\zeta_\pm$, both results show an opposite behavior, i.e., anti-correlation (see the right column). Because of the anti-correlation between $I_A$ and $I_D$ as well as $I_B$ and $I_C$, the maxima occur at $\varphi = n\pi$. Due to the phase correlation between $\xi$ and $\theta$, the maximum (minimum) occurs at $\xi = \frac{\pi}{4}$ ($\frac{3\pi}{4}$), where $\frac{3\pi}{4}$-based $R_{AD}$ is equivalent to that for $\xi = -\frac{\pi}{4}$ due to the doubled modulation frequency in equation (13). Thus, the macroscopic version of the proposed nonlocal correlation in Fig. 1 has two control parameters: one for MZI phase bases $\varphi, \psi \in \{0, \pi\}$ and another is for polarizer's rotation bases $\xi, \theta \in \{-\frac{\pi}{4}, \frac{\pi}{4}\}$. These relations are similar to those in quantum key distribution protocols such as orthogonal polarizations (H,V) of a single photon and its measurement bases (+,×) in BB84 [37,38], respectively (discussed elsewhere).

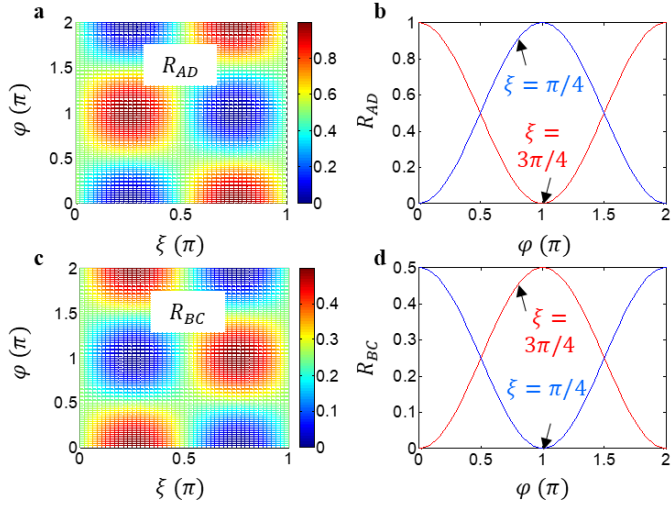

Fig. 4. Numerical simulations for cross-correlation for $\psi = 0$ and $\theta = \frac{\pi}{4}$.

**Conclusion**

Satisfying the quantum feature of "delayed choice" in a pair of orthogonally polarized MZIs composed of PBS and BS, nonlocal correlation between MZI output photons was numerically demonstrated using coherent photons via polarization-basis projections onto a set of polarizers. Unlike conventional understanding of nonlocal correlation limited to microscopic quantum particles such as entangled photon pairs, the present nonlocal scheme was for Poisson distributed coherent photons. Due to the coherence feature of the presented nonlocal system, a single photon and cw light showed the same quantum features. Like the Franson-type nonlocal correlation, the presented method is also sensitive to the phase relation between measured photons under the coincidence detection. However, this phase correlation is free from a relative distance (phase) between the photon source and two MZIs. Regarding photon indistinguishability for the nonlocal correlation, basis randomness was introduced for local randomness and applied for post-measurement-based quantum feature. Here, the basis randomness requires coherence between orthogonally polarized paired photons, where the paired photons were extended into paired cw lights, resulting in a macroscopic version of nonlocal correlation. The coincidence detection in nonlocal correlation plays a critical role for the nonlocal correlation, where the coherence feature results in much higher detection efficiency in both resolution and sensitivity.

**Acknowledgments**

This work was supported by ICT R&D program of MSIT/IITP (2021-0-01810), development of elemental technologies for ultrasecure quantum internet.